\documentclass[fleqn,twoside]{article}
\usepackage{espcrc2}

\usepackage{epsf}

\newcommand{\be}{\begin{equation}}
\newcommand{\ee}{\end{equation}}

\title{Large N reduction with overlap fermions}

\author{J. Kiskis\address[UCD]{Department of Physics, University of
California, Davis, CA 95616}
        ,
        R. Narayanan\address{Department of Physics, Florida International
University, Miami, FL 33199}
,
and
        H. Neuberger\address{Department of Physics, Rutgers University,
Piscataway, NJ 08855}\thanks{Speaker.}\thanks{Research supported in part
by the DOE under grant nr. DE-FG02-01ER41165.
}
}

\begin{document}

\begin{abstract}
We revisit quenched reduction with fermions and explain how 
some old problems can be avoided using the overlap Dirac operator.
\vspace{1pc}
\end{abstract}

\maketitle

\section{Introduction}

This contribution, based on joint work, 
provides some background to Narayanan's talk~\cite{paper}.

Lattice field theory provides an approach to calculating numerical values
or bounds for physical quantities from first principles without relying on 
perturbation theory or on non-systematic approximations. However, most
of the numerical results about strong interactions involve fermions and are
obtained in valence (quenched) QCD. This is not a systematic approximation 
because the zero flavor limit is singular. As a result, many lattice results 
still do not have a status higher than estimates coming from other 
phenomenological approximations that make ad-hoc approximations of different 
types. It makes little sense to go beyond
few percent accuracy in simulations with valence QCD. 

The long term goal of our work is to replace valence QCD with planar QCD in 
either
the 't Hooft~\cite{thooft} or the Veneziano~\cite{veneziano} 
limit. Unlike valence QCD, for many 
quantities of interest, planar
QCD is believed to be the first term in a systemtic series. Our initial focus
is on the two point functions of quark bilinears. In the planar limit, such
a correlation function is characterized by an infinite number of pole masses
and corresponding residues. Using the lattice, it should be possible to obtain
good values for these correlation functions in momentum space at Euclidean momenta
that are smaller than the inverse lattice spacing. Such information would be 
phenomenologically useful, for example, in the analysis of weak 
semi- and non-leptonic decays~\cite{derafael}. 

\section{The planar limit}
Let us consider QCD with gauge group $SU(N_c)$ instead of $SU(3)$. 
We are in Euclidean space, and assume that the volume $V$ is already
taken to infinity. Similarly, the ultraviolet cutoff $\Lambda$ has
also been eliminated. The coupling constant has disappeared too,
generating an intrinsic scale in terms of which all predictions
are pure numbers depending only on $N_c$. There are good choices for
this intrinsic scale, in the sense that they lead to simple $N_c$ dependences.
Feynman diagrams and more general counting arguments tell
us what powers of $N_c$ to scale out so that remaining
quantities have finite limits at $N_c=\infty$. In short, a nonperturbatively 
defined large $N_c$ limit seems to exist and is reachable by lattice techniques. 

In perturbation theory, the large $N_c$ limit is easiest to understand
before renormalization is taken into account. To get planar perturbative QCD
one scales the gauge coupling constant $g^2$ by a factor of $N_c$ and
$N_c$ is taken to infinity keeping $\lambda=g^2 N_c$ fixed. Using
the known terms in the $\beta$-function one easily derives how a perturbative
scale can be defined so that it stays finite at infinite $N_c$. 

The above 't Hooft limit 
generalizes to the case where the number of flavors, $N_f$ also
is taken to infinity with the ratio $\rho=\frac{N_f}{N_c}$ kept fixed.
This gives the Veneziano limit.

On the lattice, fast convergence to the $N_c=\infty$ limits was observed
for the string tension, glueball mass, topological susceptibility, and finite
temperature transition point in pure gauge theory~\cite{teper}. 
These calculations are of theoretical
interest; for practical applications it is 
certainly cheaper to just work at $N_c=3$.

Having learned that the large $N_c$ limit is meaningful outside 
perturbation theory and that it provides numbers quite close to $N_c=3$,
we note that large $N_c$ reduction~\cite{ek} provides a 
potential shortcut that would enable us to get the
planar limit numbers in a substantially more efficient way 
because of the dramatic
reduction in the number of degrees of freedom one 
needs to integrate over. In particular, the fermionic matrix becomes so small
that it can be stored in its entirety in memory and dealt
with in a more direct way than in 
standard simulations. 

\section{Quenched Reduction}
Reduction works on lattices with finite numbers of sites. In physical 
applications, reduction 
implies a change in the order of limits of the traditional
$\frac{1}{N_c}$ expansion. Now, we first take $N_c$ (possibly also $N_f$) 
to infinity and only subsequently deal with the large $V$ and 
large $\Lambda$ limits. 

The basic content of reduction is that there exists a certain ``prototype''
lattice gauge theory which encompasses lattice gauge theories with varying
group sizes on differently sized lattices. However, the entire prototype
family has the same planar limit. Thus, large $V$ can be traded for larger
$N_c$. The main point is that this tradeoff appears to be very lucrative. 

The prototype variables are $d$ $SU(N)$-matrices 
$T_\mu, ~\mu=1,2..,d$ and two
fermionic $N\times M$ complex matrices $\bar\Psi$ and $\Psi$. 
The pure gauge action is given by
\begin{equation}
S_g = \frac{N}{4\lambda} \sum_{\mu,\nu} Tr \left ( 
C_{\mu\nu}C_{\mu\nu}^\dagger \right ),~~~C_{\mu\nu}\equiv[T_\mu,T_\nu] .
\end{equation}
It is invariant under $SU(N)$ conjugation 
$T_\mu \rightarrow \Omega T_\mu \Omega^\dagger$
and under a $Z(N)^d$ acting by 
$T_\mu\rightarrow e^{\frac{2\pi\imath}{N} k_\mu} T_\mu$.

To add fermions one introduces a ``Dirac'' operator $D(T)$, which
also transforms by conjugation, and a term $S_f=Tr\left ( \bar\Psi D(T) \Psi \right )$
This breaks the $Z(N)^d$. There is an additional fermionic symmetry
under the obvious action of $SU(M)$. 

Assuming that $N$ and $M$ factorize, 
$N=nL_1 L_2 ...L_d$ and $M=mL_1 L_2 ...L_d$,  
one can constrain
the basic variables in the prototype so that one obtains a traditional 
lattice gauge theory on a toroidal $L_1\times L_2 ..\times L_d$ lattice. 
A planar limit can be defined on the lattice by scaling the lattice gauge
coupling with $n$ in the 't Hooft fashion. The basic claim of~\cite{ek} was that
for a certain class of interesting observables, the 
differently constained prototypes
become identical in this planar limit, at finite lattice spacing. 
However when the continuum limit is approached, the identity is 
spoiled by a phase transition. Quenched reduction~\cite{bhn} is designed to 
eliminate 
this effect by applying another constraint to the
prototype theory: 
The eigenvalues of the $T_\mu$ matrices are frozen at $d$ independently
drawn sets of uniformly distributed points on the unit circle parametrized 
by angles $\theta_\mu^j$. Further~\cite{ln}, the $T_\mu$ matrices entering $D(T)$
are multiplied by $U(1)$ phase factors of the form $e^{\imath p_\mu}$.
Observables obtained by carrying out a quenched average with fixed angles
have to be subsequently averaged over uniformly distributed
$\theta_\mu^j$ and $p_\mu$. A large $N$ limit can be defined as before, 
scaling the lattice coupling with $N$. 

\section{The resolution of some old problems}

In the past, when chirality and the lattice were thought to be
irreconcilable, the preferred way to introduce fermions was in
a continuum version of quenched reduction~\cite{gk}. 
In this version one replaces the unitary $T_\mu$ matrices 
of the prototype by hermitian
matrices: $T_\mu\rightarrow 1+\imath A_\mu$. 
$a_\mu^j$, the eigenvalues of $A_\mu$, are frozen and drawn from
a uniform distribution over an interval $[-\frac{\Lambda}{2},\frac{\Lambda}{2}]$.
Similarly, the Dirac operator $D(T)$ gets replaced by a $D(A)$ and the random
variable $p_\mu$ is introduced by replacing $A$ by $A_\mu+q_\mu$ with
$q_\mu$ randomly drawn again from $[-\frac{\Lambda}{2},\frac{\Lambda}{2}]$.

In perturbation theory this looks fine. Surprisingly, 
there seems to be no clash between the 
UV regularization and gauge invariance even when
the continuum theory would be chiral and anomalous. In addition, there seems
to be no room in even $d=2k$ for topological charge, since the natural object
\begin{eqnarray}
&\epsilon_{\mu_1 \nu_1 \mu_2 \nu_2....\mu_k\nu_k} \cr
&Tr \left (
[A_{\mu_1},A_{\nu_1}] [A_{\mu_2},A_{\nu_2}] ....[A_{\mu_k},A_{\nu_k}] \right )
\end{eqnarray}
vanishes identically. 

In the past, it was unclear how this regulated continuum reduced model differed
from the lattice version. It made some 
sense that topology would become a strange
concept because the disappearance of spacetime took away the concept of a local
topological density. Also, so long as fermions were dealt with in the 't Hooft limit,
the issue of anomalies did not look pressing. Nevertheless, there was enough 
uncertainty surrounding this issue that no 
further work on fermions in reduced models 
was done. The lattice provided no attractive alternative.

The main new observation in this context is that choosing $D(T)$ as the overlap
Dirac operator~\cite{overlap} frees the lattice version with fermions from all difficulties.
The overlap also provides a definition of topological 
charge~\cite{topology} in the pure gauge
reduced model, opening the possibility to be consistent with recent 
studies of topology on the lattice at large $N_c$ in traditional formulations. 
Moreover, topological obstructions that reflect anomalies are known to exist
in ordinary lattice gauge theories already on finite lattices, as emphasized
in~\cite{geom}. These obstructions will also appear in the reduced model.
Thus, so long as we stay on the lattice, quenched 
reduction will not eradicate anomalies.

\section{Comments}

Reduction seems to be closely related to the technique of ``field orbifolding''
~\cite{bersh}. For a pure $U(N_c )$ gauge theory, 
the constraints that turn 
the prototype model into a regular lattice gauge theory
are equivalent to 
imposing invariance under a subgroup of the symmetry group of the prototype. 
It is known that field orbifolding preserves the
planar limit in certain continuum filed theories 
and has a geometrical interpretation in string theory.
Could one gain new insights into reduction from this?

Numerically, twisted 
reduction~\cite{tony} used to be
considered superior to quenching in the pure gauge sector. Unfortunately, the introduction of
fermions is problematic~\cite{das}. Can one get around the fermion problem in some way
while sticking to twisted reduction instead of quenched reduction in the
pure gauge sector?

\section {Conclusions}
This is a good time to numerically revisit quenched reduction. Some of the old
problems have been solved: 
Instanton effects and anomalies can be incorporated. 
Presently available 
computational power, albeit still
short of what is needed for a full QCD simulation, 
might be amply capable of dealing with 
reduced models.


\begin{thebibliography}{9}
\bibitem{paper}  J. Kiskis, R. Narayanan and H. Neuberger, Phys. Rev. D66,
025019 (2002); talk by R. Narayanan in these proceedings.
\bibitem{thooft} G. 't Hooft, Nucl. Phys. B72, 461 (1974).
\bibitem{veneziano} G. Veneziano, Nucl. Phys. B117, 519 (1976).
\bibitem{derafael} E. de Rafael, Talk in these proceedings.
\bibitem{teper} M. Teper, hep-ph/0203203.
\bibitem{ek} T. Eguchi and H. Kawai, Phys. Rev. Lett. 48, 1063 (1982).
\bibitem{bhn} G. Bhanot, U.M. Heller and H. Neuberger, Phys. Lett. B112, 47
(1982).
\bibitem{ln} H. Levine and H. Neuberger, Phys. Lett. B119, 183 (1982).
\bibitem{gk} D. Gross and Y. Kitazawa, Nucl. Phys. B470, 369 (1982).
\bibitem{overlap} H. Neuberger, Phys. Lett. B417, 141 (1998); Phys. Lett.
B427, 353 (1998).
\bibitem{topology} R. Narayanan and H. Neuberger, 
Phys. Rev. Lett. 71, 3251 (1993); Nucl. Phys. B443, 305 (1995).
\bibitem{geom} H. Neuberger, Phys.Rev. D59, 085006 (1999).
\bibitem{bersh} M. Schmaltz, Phys. Rev. D59, 105018 (1999).
\bibitem{tony} A. Gonz\'{a}lez-Arroyo and M. Okawa, Phys. Lett, 120B, 174 (1983);
Phys. Rev. D27, 2397 (1983).
\bibitem{das} S. R. Das, Phys. Lett. B132, 155 (1983). 

\end{thebibliography}
\end{document}